CONCORDIA UNIVERSITY

# SOFTWARE REQUIREMENTS SPECIFICATION

Softbody Simulation System Design

**18/4/2013**

Prepared by:

Gustavo Barbieri Pereira – 6273033

Herman Sonfack - 5986052

Kapies Vallipuram - 9346287

Klajdi Karanxha – 6173780

# Table of Contents

Concordia University

# Vision Document

SOEN 6481

**Prepared by:**

Klajdi Karanxha **|** 6173780
Herman Sonfack **|** 5986052
Gustavo Pereira **|** 6273033
Kapies Vallipuram **|** 9346287

# Revision History

| Date | Revision # | Description | Author(s) |
|---|---|---|---|
| 01-02-2013 | 1.0 | Initial Revision of Vision Document | Klajdi Karanxha, Herman Sonfack, Kapies Vallipuram, Gustavo Pereira |
| 01-03-2013 | 2.0 | Vision Document Review/Update | Klajdi Karanxha, Herman Sonfack, Kapies Vallipuram, Gustavo Pereira |
| 17-04-2013 | 3.0 | Final Review | Klajdi Karanxha, Herman Sonfack, Kapies Vallipuram, Gustavo Pereira |

# Table of Contents

# 1. Introduction
## 1.1 Document purpose and scope

The purpose of this document is to give the vision for interactive computer graphics physical based simulation systems. It focuses on the needs of stakeholders and the reasons for such needs.

The vision document is directed toward the Softbody Simulation System. This serves as a typical physical-based simulation system involving real time computer graphics. Understanding this system and its requirements will help to understand similar systems and the requirements they need to have. Softbody Simulation System is academic, open-source and extendable to easily adapt to new requirements. It visualizes the simulation process using libraries and applications. The system is interactive and allows altering the simulation at run-time. The level-of-detail and simulation parameters can also be manipulated. The system has other features, such as vertex and fragment shader program and support in the cross-vendor GPU assembly language, as well as the OpenGL Shading Language (GLSL).

## 1.2 Reference

"Deriving Software Engineering Requirements Specification for Computer Graphics Simulation Systems through a Case Study" by Miao Song and Peter Grogono, Concordia University.

# 2. Positioning
## 2.1 Problem Statement

| | |
|---|---|
| **The problem of** | No reference when specifying and designing new Computer graphics systems. |
| **Affects** | Researchers, software architects, teachers. |
| **The impact of which is** | Interactive simulation where users can modify parameters at run time. |
| **A successful solution would be** | Open-source and extendable system that visualizes the simulation process using OpenGL and other related Libraries and applications. It is an interactive system that allows the user to alter the simulation at run-time via mouse and keyboard. It also includes a GLUI-based graphical controls to manipulate the level-of-detail (LOD) and simulation parameters. The system has other features, such as vertex and fragment shader program, support in the cross-vendor GPU assembly language as well as the OpenGL Shading Language. |

## 2.2 Product Position Statement

| For | Researchers |
|---|---|
| Who | Test different physical based phenomena at real time. |
| The Softbody simulation | Is a software product. |
| That | Allow real time interactive simulation. |
| Unlike | Other commercial simulation and visualization software packages, such as Fluent for fluid dynamics and its Gambit (both usually work together for simulation and later visualization or large open-source projects that can be used for rendering various computer graphics-related materials in simulation, such as OGRE. |
| Our product | Is an integrated solution, open source and in continuous development for more features. |

# 3. Stakeholder Descriptions

## 3.1 Stakeholder Summary

| Name | Description | Responsibilities |
|---|---|---|
| Researchers | The users of the product. | Test different physical based phenomena at real time. |
| Software architects | People that develop similar systems | Specifying and designing such kind of computer graphics systems. |

## 3.2 User Environment

The system compiles and runs on the Microsoft Windows platform, but with growth it is natural to accommodate other platforms, such as Linux and Mac OS X. The Softbody Simulation System is written in C++, and is source-code portable. The user's environment should have their video drivers up to date in order to handle the latest version of Open GL. Hardware specs are not entirely clear yet, but will require minimal specifications to handle the computation to render complex visuals.

# 4. Product Overview

This section provides a high-level view of the product capabilities, interfaces to other applications.

## 4.1. Product Perspective

The Softbody Simulation System is an interactive system that uses OpenGL and other related libraries and applications for visualization of the simulation process.

It will allow enhancing on interactivity by external devices/applications, as re-active controls that include haptic devices to provide force feedback. Such applications can be used e.g. for surgeon training, games, or even interactive cinema.

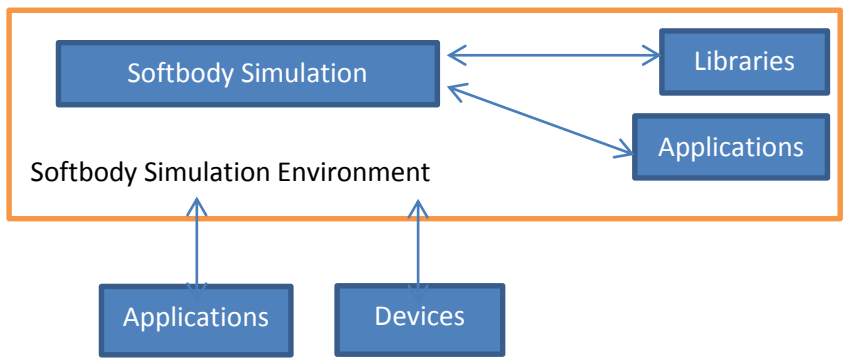

## 4.2. Assumptions and Dependencies

| Assumptions | Dependencies |
|---|---|
| Server is running compatible OS | Application runs on Windows, Linux and Mac OS X environments |
| Server support OpenGL | System compatible with latest versions of OpenGL, updated drivers (Windows OS) or GLX extensions (Linux OS) required |
| System has academic value of teaching and learning computer graphics / physical based simulations | Structured source-code, comments and documentation per consistent naming and coding conventions |
| Portability of the source code | Export tool and generation of the system as a library or a collection of libraries and APIs |

## 4.3. Needs and Features

| Need | Priority | Features | Planned Release |
|---|---|---|---|
| Implement statistics gathering for various real-time performance metrics for Simulation and Rendering | Low | - Simulation Performance Report<br>- Rendering Performance Report | Iteration 3 |
| Allow the user to alter the simulation at run-time via GLUI-based graphical controls to manipulate the level-of-detail (LOD) and simulation parameters | High | User I/O LOD GUI | Iteration 1 |
| Run-time selection for comparative studies on any aspect of visual realism to run-time performance and memory usage | Normal | Visual Performance Analysis | Iteration 2 |
| Lighting and texture mapping techniques | High | Lighting&Texture | Iteration 1 |
| Allow objects be "attached" to "hardbody" objects providing points of attachment | High | Attachment | Iteration 1 |
| Allow alteration of the Archimedean-based graphs and different types of them than a octahedron as a run-time LOD parameter | Normal | Archimedean-based graphs Alteration | Iteration 2 |
| Interactivity through haptic devices | Low | Device Control | Iteration 3 |
| Allow the state dump and reload functionality at any given point in time enabling to reproduce a simulation from some point in time | Normal | Run Time Watch Point | Iteration 2 |
| Allow for stereoscopic effects | Low | Stereoscopic Effects | Iteration 3 |

## 4.4. Alternatives and Competition

There are many commercial simulation and visualization software packages, as Fluent for fluid dynamics and Gambit by ANSYS and Matlab's visualization. None of them are open-source, what implies in huge investments. They have biggest size and amount of man-power and resources involved, but have system-domain-specific requirements and they don't have portability. Large open-source projects can be used for rendering various computer graphics-related materials in simulation, such as OGRE.

# 5. Other Product Requirements

## 5.1 Platform requirements

We plan to compile and run the application on a Windows, Linux and Mac OS X environment. The client requires having a computer with any of these operating systems. Since OpenGL will be used for the graphics, it is portable throughout all the different environments. For the Linux environment we will require use of GLX extensions to the X Server. On the windows environment you will be required to have the drivers to support latest versions of OpenGL.

## 5.2 Hardware requirements

Video card has drivers to support latest versions of Open GL and has at least 256 MB of RAM.

## 5.3 Performance requirements

We would like to be able to reduce the load on run time, therefore plan for allowance for multiple algorithms run-time selection for comparative studies on any aspect of visual realism to run-time performance and memory usage. Be able to support multiple displays without slowing down the system. ALL I/O devices should take into account level of detail which will be handled by the algorithms to avoid unnecessary computations yet delivering adequate visual quality.

## 5.4 Usability requirements

Usability isn't a huge concern at the moment but there should be an importance to test different physical based phenomena at real-time.

## 5.5 Standards

Take into consideration the academic value of teaching and learning computer graphics and physical based simulations. This will be done by structuring the code, comments and documentation per consistent naming and coding conventions. This will be highly important for open-source and academic projects.

## 5.6 Documentation

Documentation will be carried out carefully because the project will be used as an open source project and will be used to academic use. Therefore every aspect will be carefully documented with proper code standards. This will result in artifacts such as a user manuals and installation documents.

# Appendix A

## Interview

Christian Blais is Systems Architect at CAE Inc. The department he works for delivers training to a specific audience OD pilots using graphical simulation tools in class or independently. Recently, they want to discard all the many graphic simulation tools they use to deliver their training in favor of a single tool that would be able to perform as a sum of all existing ones. They also want the source code of this tool so they could be able to understand, maintain and scale it in the future into a proprietary technology used specifically for their simulation needs.

## Part I – User Profile

**Name:** Christian Blais

**Company:** CAE Inc.

**Industry:** Aerospace and Simulation

**Job Title:** Systems Architect

### What are your key responsibilities?

Converting problem solutions into technological solutions and mapping requirements into features. Finding technology solutions to problems and contributing in existing products innovation. Design development and occasionally direct development effort.

### What outputs do you produce?

Technical solutions to existing problems, Software architecture documentation and source code during development phase.

### For whom?

Interested stakeholders and management team.

### How is success measured?

Management approval of suggested technical solutions as well as customer feedback and satisfaction.

### Which problems interfere with your success?

Unrealistic deadlines and expectations sometimes create an issue. In addition, difficulty to clearly convey solution benefits to management team creates an issue.

### What, if any, trends make your job easier or more difficult?

Change in requirements and unstable requirements usually inflict difficulty. Mental model and system visualization from clients sometimes does not match and is inconsistent with the system being used or being improved here.

## Part II - Assessing the Problem

### For which Graphic Simulation problems do you lack good solutions? What are they?

We have many graphic simulation tools and systems to deliver our courseware to our clients. However we lack good solutions on this aspect as we would rather unify our tools and go with a system that is easily supported and continuously improved. In terms of poor graphic simulation solutions we should need a tool that offers more support, requires minimum configuration and adapts to client's changing needs.

*Problem:* Lack of a unification tool to deliver courseware related graphic solutions to clients.

### Why does this problem exist?

Courseware delivery and courseware content changed over time with increasing needs in the aerospace market. We need more realistic graphic simulation tools to provide training to our clients. With changing demands we adapted new tools in addition to existing ones.

### How do you solve it now?

Presently, we have many graphic simulation tools that can solve different parts of the same problem and that are what we use.

### How would you like to solve it?

Adapt a unitary graphic simulation tool which we can build on and scale it over time with growing demand, rather than adapting many tools.

## Part III - Understanding the User Environment

### Who are the users?

Clients, Software Architects and developers.

### What is their educational background?

Usually university or college equivalent degree.

### What is their computer background?
It ranges from medium skills to tech experts.

### Are users experienced with this type of application?
That depends on user type. Clients, we usually do not expect them to be familiar or experienced with this. Developers, who will contribute, should be familiar with main concepts.

### Which platforms are in use?
Windows 7, MAC OS X 10.6

### What are your plans for future platforms?
Support Linux platforms as well and mobile platforms such as Android or iOS.

### Are additional applications in use that is relevant to this application?
Not for the moment.

### What are your expectations for usability of the product?
We have very high usability expectations given the sensitivity of training we offer.

### What are your expectations for training time?
As little as possible for developers and working out of the box for clients.

### What kinds of user help do you need?
We need online documentation.

## Part IV – Recap for Understanding
*You have told me:*

6. Your company delivers training to a specific audience of customers. Training is complex and requires graphic simulation solutions.
7. Your current graphic solutions are decentralized, per demand basis and do not fully solve your problem.
8. You would like to unify the graphic simulation tools into a single unitary tool that you will maintain, improve and scale based on customer demand.

### Does this adequately represent the problem you are having with your existing solution?
Yes, it does.

## What, if any, other problems are you experiencing?
The above mentioned is the only one for the moment.

# Part V - The Analyst's Inputs on the Customer's Problem
## Which, if any, problems are associated with:
- Lack of a robust, unifying graphic simulation tool.

## Is this a real problem?
Not having a single graphic simulation tool is a significant issue at our department as maintenance cost of other legacy solutions is becoming higher and also in order to maintain our clients we have to be always ahead of competition.

## What are the reasons for this problem?
There are many reasons associated to this issue, precisely: changes in management decisions, inability to keep up and lack of inter-department communication. Also, inability to capture 'voice of customer' in my opinion.

## How do you currently solve the problem?
Currently, as stated, we have different graphic simulation tools that solve different parts of the same problem.

## How would you like to solve the problem?
As mentioned, we would like a single graphic simulation tool that we will maintain and use.

## How would you rank solving these problems in comparison to others you've mentioned?
I would rank solving this problem as high priority.

# Part VI - Assessing Your Solution
## What if you could?
- Have access to a system solution which is based on open source components but it is designed to be scaled and performing.
- Have access to a system solution that is designed following an architecture that is adapted to graphic development.
- Have access to a system solution that is designed to scale in terms of platform support.
- Have access to a system solution that is based on a technology fundamental to flight simulation, video games and offers an API to interact as needed with the GPU.
- Have access to a system solution that once translated to a product requires minimum user

configuration and works out of the box.

## How would you rank the importance of these?
1. Have access to a system solution that once translated to a product requires minimum user configuration and works out of the box.
2. Have access to a system solution that is based on a technology fundamental to flight simulation, video games and offers an API to interact as needed with the GPU.
3. Have access to a system solution that is designed following an architecture that is adapted to graphic development.
4. Have access to a system solution which is based on open source components but it is designed to be scaled and performing.
5. Have access to a system solution that is designed to scale in terms of platform support.

# Part VII - Assessing the Opportunity
## Who in your organization needs this application?
Management to showcase to client. System architect to suggest solution.

## How many of these types of users would use the application?
The number would probably be in hundreds.

## How would you value a successful solution?
If there is positive feedback from the clients and quick familiarization from the developers that would be a valued solution.

# Part VIII - Assessing the Reliability, Performance, and Support Needs
## What are your expectations for reliability?
Very High. The quality of our work relies on this solution.

## What are your expectations for performance?
Normal. A light solution with no visual delays (lagging) is preferred.

## Will you support the product, or will others support it?
We will support it.

## Do you have special needs for support?
Yes. We need to maintain our own solution documentation.

### What about maintenance and service access?
We will maintain the solution once we own it.

### What are the security requirements?
Cannot be identified at the moment.

### What are the installation and configuration requirements?
It is essential to work on existing configurations; support for other platforms is preferred. Currently, our graphical solutions can be run from many ordinary computers (Win7, Intel-based architecture and low end graphic card or chipset.)

### Are there special licensing requirements?
Once we own the solution whatever modification we perform on it will be proprietary and restricted under our own associated policy.

### How will the application be distributed?
Through online downloads. Electronically.

### Are there labeling and packaging requirements?
Yes, our company branding and labeling policy will be applied.

## Part IX - Other Requirements
### Are there any legal, regulatory, or environmental requirements or other standards that must be supported?
Yes, since part of our company is in business with U.S Department of State, International Traffic in Arms Regulations will regulate the product development and distribution.

### Can you think of any other requirements we should know about?
Not for now.

## Part X - Wrap-up
### Are there any other questions I should be asking you?
No

### If I need to ask follow-up questions, may I give you a call? Would you be willing to participate in a requirements review?
Of course.

# Part XI - The Analyst's Summary

1. A unitary graphic simulation solution to deliver training to clients with good performance on normal usage computers.
2. A scalable solution based on industry standards which requires minimal training and will be evolved by the customer.
3. A graphical simulation solution that would allow greater interoperability in other platforms.

# Glossary

**API** – Application Program Interface is a protocol intended to be used as an interface by software components to communicate with each other. [1]

**Chipset** – Usually refers to a group of semiconductor chips which perform a specialized function, in this case the functions of a graphic card.

**GPU** - The GPU (Graphics Processing Unit) is a specialized circuit designed to accelerate the image output in a frame buffer intended for output to a display.

**ITAR** – International Traffic in Arms Regulations is a set of legal regulations issued by the U.S Department of State which applies to every company that is in business with the U.S Military and as stated in Section 38 of the Arms Export Control Act (22 U.S.C. 2778) authorizes the President (U.S President) to control the export and import of defense articles and defense services.[3]

**Open Source** – Is a software development or broader philosophy which promotes free distribution of software design and implementation.

**Out-of-the-Box** - A feature / product that is intended to work without major end user intervention.

**Performance** - the execution of an action [4]

**Platform** – A computing platform usually refers to the hardware architecture and the software framework of a computer system.[7]

**Proprietary** – Refers to computer software or any other kind of solution which is licensed and used or delivered exclusively by the copyright holding company or individual.[6]

**Source Code** – In computer science, source code is any collection of computer instructions (possibly with comments) written using some human-readable computer language, usually as text. The source code of a program is specially designed to facilitate the work of computer programmers, [5]

Concordia University

# Use Case Briefs

SOEN 6481

**Prepared by:**

Klajdi Karanxha **|** 6173780
Herman Sonfack **|** 5986052
Gustavo Pereira **|** 6273033
Kapies Vallipuram **|** 9346287

# Revision History

| Date | Revision # | Description | Author(s) |
|---|---|---|---|
| 01-03-2013 | 1.0 | Use Case Briefs | Klajdi Karanxha, Herman Sonfack, Kapies Vallipuram, Gustavo Pereira |
| 01-04-2013 | 2.0 | Passive Voice, Weak Phrases optimization | Klajdi Karanxha, Herman Sonfack, Kapies Vallipuram, Gustavo Pereira |
| 17-04-2013 | 3.0 | Final Review | Klajdi Karanxha, Herman Sonfack, Kapies Vallipuram, Gustavo Pereira |

# Introduction

### 1.1 Use Case 1: Create simulation Object

### Actor: Researcher

The researcher creates the elastic object with particle, springs and faces. The researcher specifies the dimensionality, one-, two-, or three-dimensional and an integrator. The object can be either imported from file or create in the live environment .The object is displayed in the view space.

### 1.2 Use Case 2: Interact with the system

### Actor: Researcher

The researcher will interact with the system using the mouse or the GLUI panel for a better level of detail (LOD). While the simulation is running the user has the option to either use the mouse or expand a box of controls that he can use to alter the simulation. The user expands the graphic control toolbar and selects any one of the controls. This prompts a new window to pop up with various changeable values belonging to a certain property of the simulation. As the user alters the values the simulation will adjust accordingly. The user can select a specific or multiple algorithms and confirm the selection. The researcher will take the statistics for each algorithm. The researcher will carry out comparative studies on any aspects of visual realism to run-time performance and memory usage.

### 1.3 Use Case 3: Exchange Simulation Data

### Actor: Researcher

The user will export information from the simulation. He could either take a state dump of the system in order to display each particle and spring state (all the force contributions, velocity, and the position) at any given point in time in a text or XML file for further import into a relational database or an Excel spreadsheet for plotting and number analysis, or record the simulation by taking the state of the system at different point of time.

### 1.4 Use Case 5: Run in idle

### Actor: System

The system runs continuously and at every given time DT, elastic objects tells the system how the objects behave and the change for their velocity and position

## 1.5 Use Case 6: Interact with re-active controls

### Actor: System

The system interacts with re-active controls that include haptic devices to provide force feedback. The soft body -like object could also be able to be attached to hard body objects and provide points of attachment.

Concordia University

# Cost-Value Prioritization & AHP Method

SOEN 6481

**Prepared by:**

Klajdi Karanxha | 6173780
Herman Sonfack | 5986052
Gustavo Pereira | 6273033
Kapies Vallipuram | 9346287

## Revision History

| Date | Revision # | Description | Author(s) |
|---|---|---|---|
| 01-03-2013 | 1.0 | Cost-Value Priorization & AHP Method | Klajdi Karanxha, Herman Sonfack, Kapies Vallipuram, Gustavo Pereira |
| 17-04-2013 | 2.0 | Final Review | Klajdi Karanxha, Herman Sonfack, Kapies Vallipuram, Gustavo Pereira |

# Table of Contents

# 1. Cost-Value prioritization

**Requirement #1**: Allow the user to alter the simulation at run-time via GLUI-based graphical controls to manipulate the level-of-detail (LOD) and simulation parameters

**Requirement #2**: Implement statistics gathering for various real-time performance metrics for Simulation and Rendering

**Requirement #3**: Run-time selection of different algorithms for comparative studies on any aspect of visual realism to run-time performance and memory usage

**Requirement #4**: Lighting and texture mapping techniques

**Requirement #5**: Allow objects be "attached" to "hardbody" objects providing points of attachment

**Requirement #6**: Allow alteration of the Archimedean-based graphs and different types of them than a octahedron as a run-time LOD parameter

**Requirement #7**: Interactivity through haptic devices

**Requirement #8**: Allow the state dump and reload functionality at any given point in time enabling to reproduce a simulation from some point in time

**Requirement #9**: Allow for stereoscopic effects

Scale for pairwise comparison

| Relative intensity | Definition | Explanation |
| --- | --- | --- |
| 1 | Of equal value | Two requirements are of equal value. |
| 3 | Slightly more value | Experience slightly favors one requirement over another. |
| 5 | Essential or strong value | Experience strongly favors one requirement over another. |
| 7 | Very strong value | A requirement is strongly favored and its dominance is demonstrated in practice. |
| 9 | Extreme value | The evidence favoring one over another is of the highest possible order of affirmation |
| 2, 4, 6, 8 | Intermediate values between two adjacent judgments | When compromise is needed. |

### a. Criteria: Value

#### i. Step 1: Normal

|      | Req1 | Req2 | Req3 | Req4 | Req5 | Req6 | Req7 | Req8 | Req9 |
|------|------|------|------|------|------|------|------|------|------|
| **Req1** | 1    | 3    | 3    | 5    | 5    | 7    | 7    | 7    | 9    |
| **Req2** | 1/3  | 1    | 1    | 3    | 3    | 5    | 5    | 5    | 7    |
| **Req3** | 1/3  | 1    | 1    | 3    | 3    | 5    | 5    | 5    | 7    |
| **Req4** | 1/5  | 1/3  | 1/3  | 1    | 1    | 3    | 3    | 3    | 5    |
| **Req5** | 1/5  | 1/3  | 1/3  | 1    | 1    | 3    | 3    | 3    | 5    |
| **Req6** | 1/7  | 1/5  | 1/5  | 1/3  | 1/3  | 1    | 1    | 1    | 3    |
| **Req7** | 1/7  | 1/5  | 1/5  | 1/3  | 1/3  | 1    | 1    | 1    | 3    |
| **Req8** | 1/7  | 1/5  | 1/5  | 1/3  | 1/3  | 1    | 1    | 1    | 3    |
| **Req9** | 1/9  | 1/7  | 1/7  | 1/5  | 1/5  | 1/3  | 1/3  | 1/3  | 1    |

#### ii. Step 2: Normalized

|      | Req1 | Req2 | Req3 | Req4 | Req5 | Req6 | Req7 | Req8 | Req9 | Relative Value |
|------|------|------|------|------|------|------|------|------|------|----------------|
| **Req1** | 0.38 | 0.47 | 0.47 | 0.35 | 0.35 | 0.27 | 0.27 | 0.27 | 0.21 | **0.34** |
| **Req2** | 0.13 | 0.16 | 0.16 | 0.21 | 0.21 | 0.19 | 0.19 | 0.19 | 0.16 | **0.18** |
| **Req3** | 0.13 | 0.16 | 0.16 | 0.21 | 0.21 | 0.19 | 0.19 | 0.19 | 0.16 | **0.18** |
| **Req4** | 0.08 | 0.05 | 0.05 | 0.07 | 0.07 | 0.11 | 0.11 | 0.11 | 0.12 | **0.09** |
| **Req5** | 0.08 | 0.05 | 0.05 | 0.07 | 0.07 | 0.11 | 0.11 | 0.11 | 0.12 | **0.09** |
| **Req6** | 0.05 | 0.03 | 0.03 | 0.02 | 0.02 | 0.04 | 0.04 | 0.04 | 0.07 | **0.04** |
| **Req7** | 0.05 | 0.03 | 0.03 | 0.02 | 0.02 | 0.04 | 0.04 | 0.04 | 0.07 | **0.04** |
| **Req8** | 0.05 | 0.03 | 0.03 | 0.02 | 0.02 | 0.04 | 0.04 | 0.04 | 0.07 | **0.04** |
| **Req9** | 0.04 | 0.02 | 0.02 | 0.01 | 0.01 | 0.01 | 0.01 | 0.01 | 0.02 | **0.02** |

### b. Criteria: Cost

#### i. Step 1: Normal

|      | Req1 | Req2 | Req3 | Req4 | Req5 | Req6 | Req7 | Req8 | Req9 |
|------|------|------|------|------|------|------|------|------|------|
| **Req1** | 1    | 3    | 3    | 5    | 1    | 1/3  | 1/3  | 5    | 5    |
| **Req2** | 1/3  | 1    | 1    | 3    | 1/3  | 1/7  | 1/7  | 3    | 3    |
| **Req3** | 1/3  | 1    | 1    | 3    | 1/3  | 1/7  | 1/7  | 3    | 3    |
| **Req4** | 1/5  | 1/3  | 1/3  | 1    | 1/5  | 3    | 3    | 1    | 1    |
| **Req5** | 1    | 3    | 3    | 5    | 1    | 1/3  | 1/3  | 5    | 5    |
| **Req6** | 3    | 7    | 7    | 1/3  | 3    | 1    | 1    | 1/3  | 1/3  |
| **Req7** | 3    | 7    | 7    | 1/3  | 3    | 1    | 1    | 1/3  | 1/3  |
| **Req8** | 1/5  | 1/3  | 1/3  | 1    | 1/5  | 3    | 3    | 1    | 1    |
| **Req9** | 1/5  | 1/3  | 1/3  | 1    | 1/5  | 3    | 3    | 1    | 1    |

### ii. Step 2: Normalized

|      | Req1 | Req2 | Req3 | Req4 | Req5 | Req6 | Req7 | Req8 | Req9 | Relative Value |
|------|------|------|------|------|------|------|------|------|------|----------------|
| **Req1** | 0.11 | 0.13 | 0.13 | 0.25 | 0.11 | 0.03 | 0.03 | 0.25 | 0.25 | **0.14** |
| **Req2** | 0.04 | 0.04 | 0.04 | 0.15 | 0.04 | 0.01 | 0.01 | 0.15 | 0.15 | **0.07** |
| **Req3** | 0.04 | 0.04 | 0.04 | 0.15 | 0.04 | 0.01 | 0.01 | 0.15 | 0.15 | **0.07** |
| **Req4** | 0.02 | 0.01 | 0.01 | 0.05 | 0.02 | 0.25 | 0.25 | 0.05 | 0.05 | **0.08** |
| **Req5** | 0.11 | 0.13 | 0.13 | 0.25 | 0.11 | 0.03 | 0.03 | 0.25 | 0.25 | **0.14** |
| **Req6** | 0.32 | 0.30 | 0.30 | 0.02 | 0.32 | 0.08 | 0.08 | 0.02 | 0.02 | **0.16** |
| **Req7** | 0.32 | 0.30 | 0.30 | 0.02 | 0.32 | 0.08 | 0.08 | 0.02 | 0.02 | **0.16** |
| **Req8** | 0.02 | 0.01 | 0.01 | 0.05 | 0.02 | 0.25 | 0.25 | 0.05 | 0.05 | **0.08** |
| **Req9** | 0.02 | 0.01 | 0.01 | 0.05 | 0.02 | 0.25 | 0.25 | 0.05 | 0.05 | **0.08** |

## 2. Diagrams

### a. Value Diagram

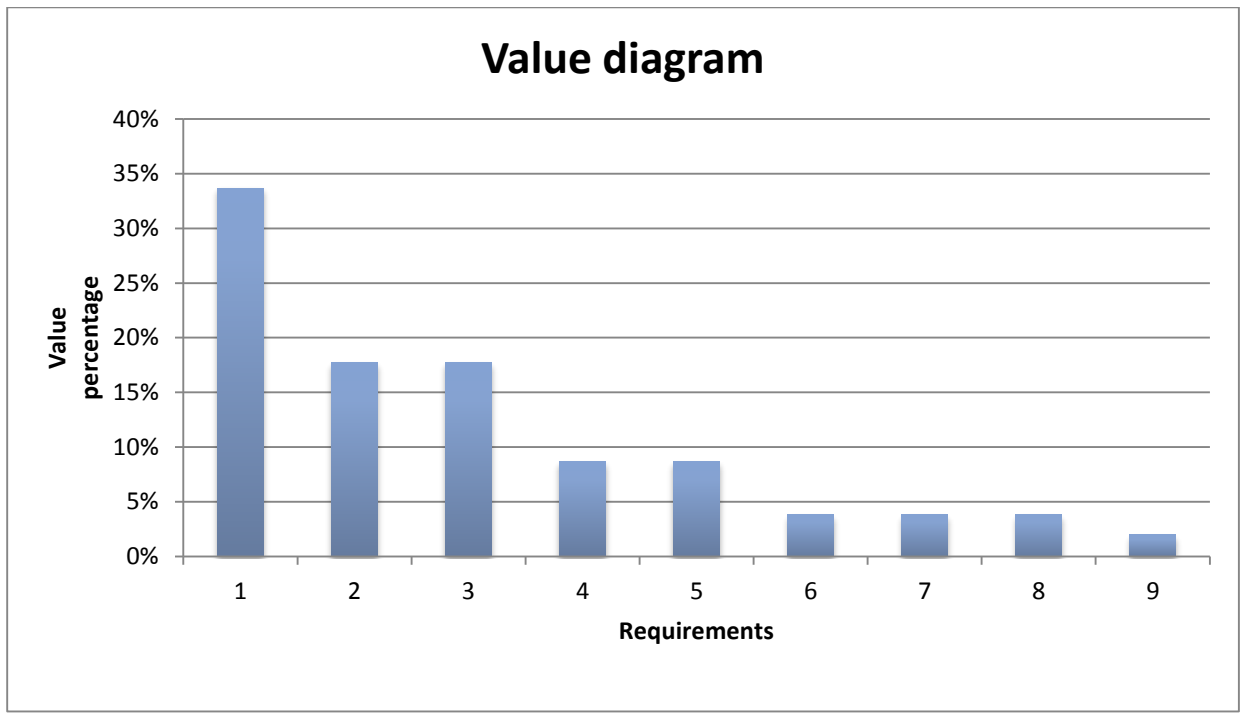

b. Cost Diagram

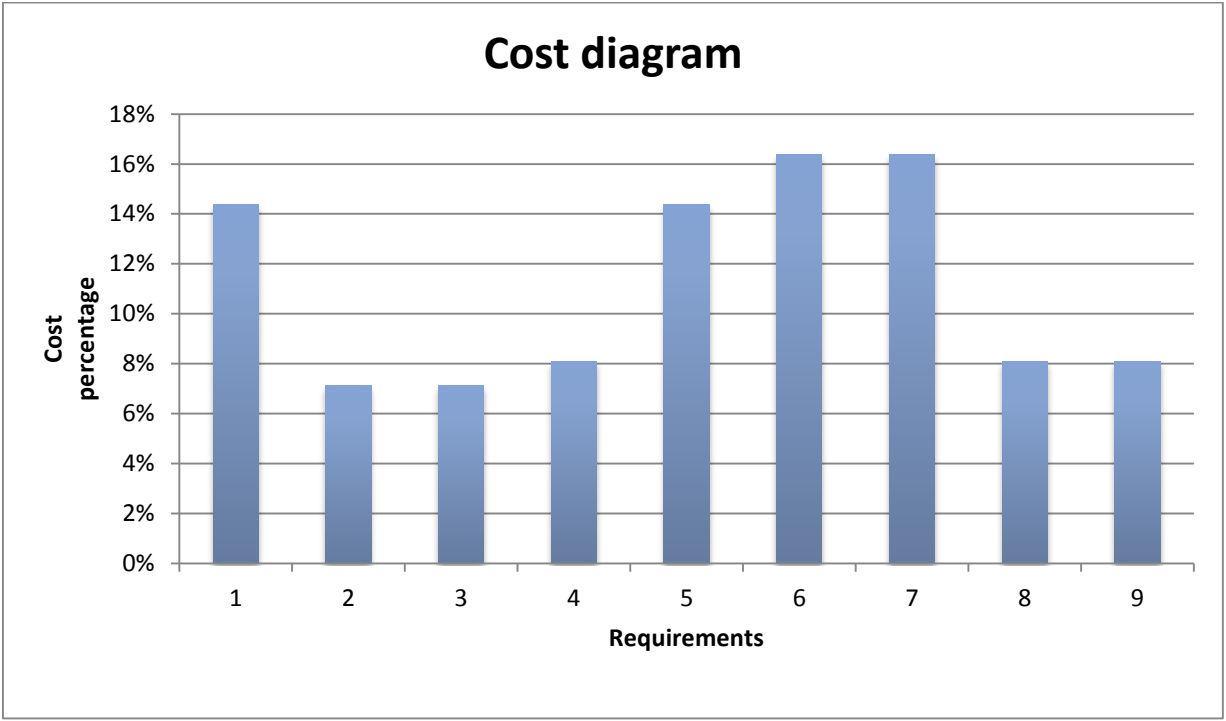

## c. Cost-Value Chart

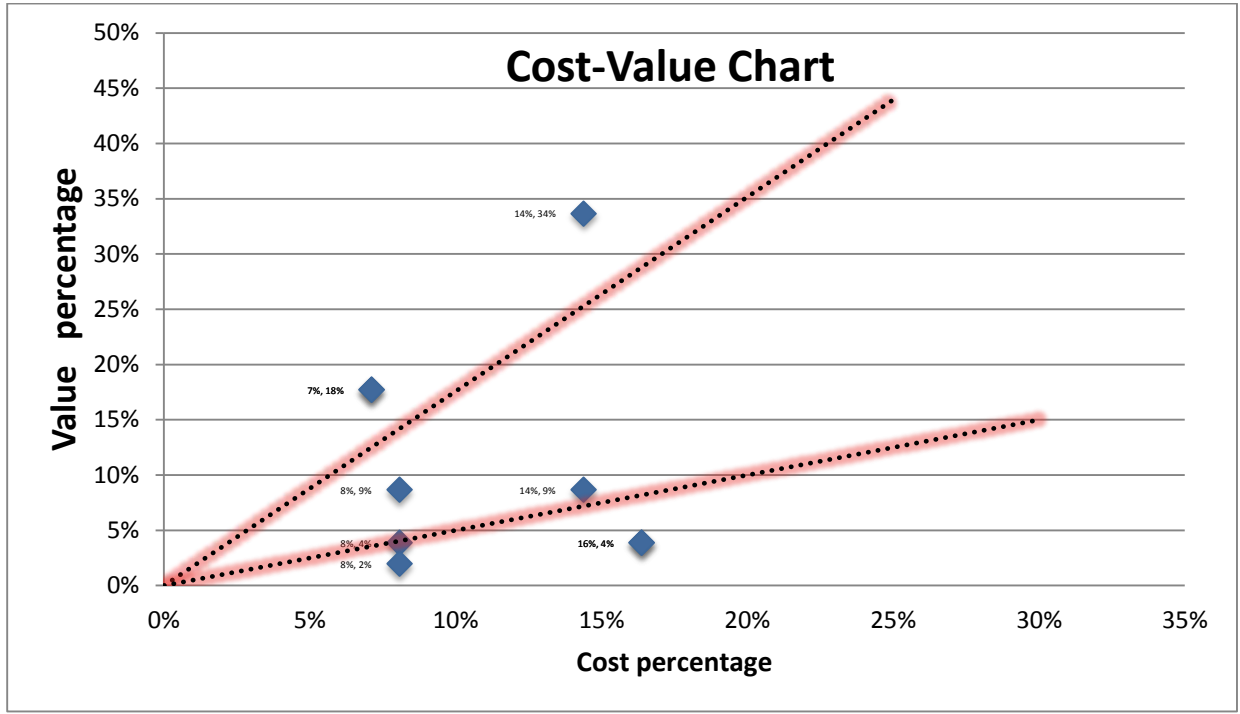

 Two cases we will select are Requirements 1 which will take 14% of the cost and has a business value of 34%. The other requirement can be requirement 2 or 3 which both take about 7% cost and give a business value of 18% to the stakeholders.

Concordia University

# Fully Dressed Use Cases

SOEN 6481

**Prepared by:**

Klajdi Karanxha **|** 6173780
Herman Sonfack **|** 5986052
Gustavo Pereira **|** 6273033
Kapies Vallipuram **|** 9346287

## Revision History

| Date | Revision # | Description | Author(s) |
|---|---|---|---|
| 01-03-2013 | 1.0 | Initial Document of a fully described Use Case, using Pre and Post conditions, extensions, Special Requirements and Use Case Context diagram. | Klajdi Karanxha, Herman Sonfack, Kapies Vallipuram, Gustavo Pereira |
| 01-04-2013 | 2.0 | Review for weak phrases, passive voice… | Klajdi Karanxha, Herman Sonfack, Kapies Vallipuram, Gustavo Pereira |
| 17-04-2013 | 3.0 | Final Review | Klajdi Karanxha, Herman Sonfack, Kapies Vallipuram, Gustavo Pereira |

# Table of Contents

# 1. Fully-Dressed Use Case

## 1.1 Use Case 2: Interact with the system

| | |
|---|---|
| **ID** | UC 2.0 |
| **Use Case** | Interact with the system |
| **Description** | A researcher can interact with the system at runtime to adjust the LOD and simulation parameters. |
| **Level** | User Level |
| **Primary Actor** | Researcher |
| **Stakeholder Interests** | <table><tr><th>Stakeholder</th><th>Interests</th></tr><tr><td>Researchers</td><td>Want to edit the simulation at runtime by adjusting various parameters to analyze different outputs and see how it behaves.</td></tr><tr><td>Software architects</td><td>Want the users of the system to be able to edit the simulation at runtime without having the system lag, or ruin the simulation currently running.</td></tr></table> |
| **Preconditions** | Simulation software has been setup and simulation is running. |
| **Post conditions** | Success end condition : <br> Simulation is running under the adjusted variables. <br> Failure end condition: <br> Simulation has crashed and is no longer running. <br> Minimal Guarantee : <br> Simulation is running but has not adjusted to new values changes. |
| **Main Scenario** | 1. User edits simulation parameters on toolbox. <br> 2. System updates presentation on display. <br> 3. System keeps running simulation under new parameters. |
| **Extensions** | [GLUI-based graphical controls] <br> 1a. Instead of using mouse/keyboard, user manipulate level-of-detail using GLUI-based graphical controls <br>     1. User edits simulation parameters using GLUI-based graphical controls. <br>     2. System automatic updates presentation on display. <br>     3. System keeps running simulation under new parameters. <br><br> [Select various algorithms at runtime] <br> 1b. user selects various algorithms from the list of algorithms on display and confirms selection <br>     1. User edits simulation parameters selecting multiple algorithms <br>     2. System automatic updates presentation on display. <br>     3. System keeps running simulation under new parameters. |

[System Failure]
1c. Option to edit parameters is disabled
   1. User exports state of current configurations onto working environment.
   2. User will manually reset to default parameters clicking the Reset button.
   3. User reloads previous configuration.
   4. The System runs and shows the previous configuration.
   5. User selects option to edit simulation parameters

[System Failure]
2a. System does not automatically update presentation on display.
   1. User exports state of current configurations onto working environment.
   2. User will manually reset to default parameters clicking the Reset button.
   3. User reloads previous configuration.
   4. The System runs and shows the previous configuration.
   5. User selects option to edit simulation parameters

## 1.2 Use Case 3: Exchange Simulation Data

| | |
|---|---|
| **ID** | UC 3.0 |
| **Use Case** | Exchange Simulation Data |
| **Description** | A researcher can export simulation data. A researcher can take a state dump of the system in order to display each particle and spring state (all the force contributions, velocity, and the position) at any given point in time and save it in a text file for further import. |
| **Level** | User Level |
| **Primary Actor** | Researcher |
| **Stakeholder Interests** | <table><tr><th>Stakeholder</th><th>Interests</th></tr><tr><td>Researchers</td><td>Want to use simulation information in a relational database or an Excel spreadsheet for plotting and number analysis.</td></tr><tr><td>Software architects</td><td>Analyze properties or problems that can be found on simulation data. Reload or use in an external tool for analysis.</td></tr></table> |
| **Preconditions** | Simulation software has been setup and simulation is running. |
| **Post conditions** | Success end condition :<br>A text file with a state dump of simulation data was written at the folder specified by the user.<br>Failure end condition:<br>Error in writing a file at the path specified by the user.<br>Minimal Guarantee :<br>File was written with Simulation state dump in the temporary folder. |
| **Main Scenario** | 1. User selects Exchange option under menu toolbar.<br>2. System display exchange options.<br>3. User selects option to Export Simulation state dump to text file.<br>4. The System asks the file name and folder to the user.<br>5. User select folder and file name.<br>6. System saves simulation data into temporary folder<br>7. System export text file to selected file name. |
| **Extensions** | [Record Simulation data]<br>3a. The researcher can record the simulation by taking the state of the system at different points of time.<br>   1. User selects option to Record Simulation and duration.<br>   2. The System asks the file name and folder to the user.<br>   3. User select folder and file name.<br>   4. System starts recording Simulation data.<br>   5. User select stop recording under menu toolbar.<br>   6. System starts procedure from step 6.<br><br>[Import Simulation data]<br>3b. The researcher can import simulation data as a replay. |

1. User selects option to import from text file.
2. The System asks the file name and folder to the user.
3. User select folder and file name.
4. System import text file with simulation data.
5. System runs simulation using simulation data from file.

[System Failure]
7a. The System cannot write the text file on specified folder
 1. The System asks a new the file name and folder to the user.
 2. System starts procedure from step 7.

## Context Use-Case Diagram

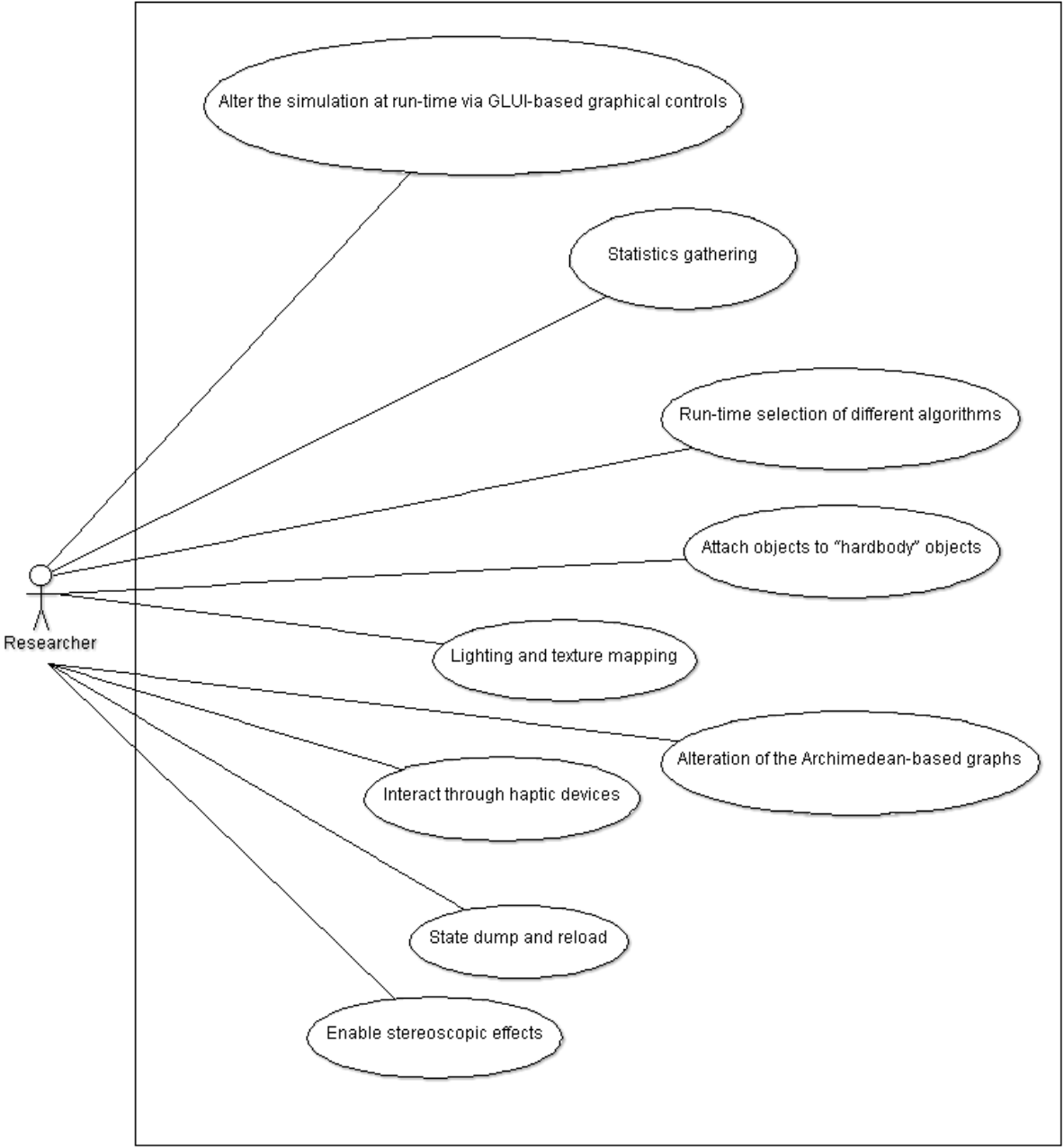

**Requirement 1:**

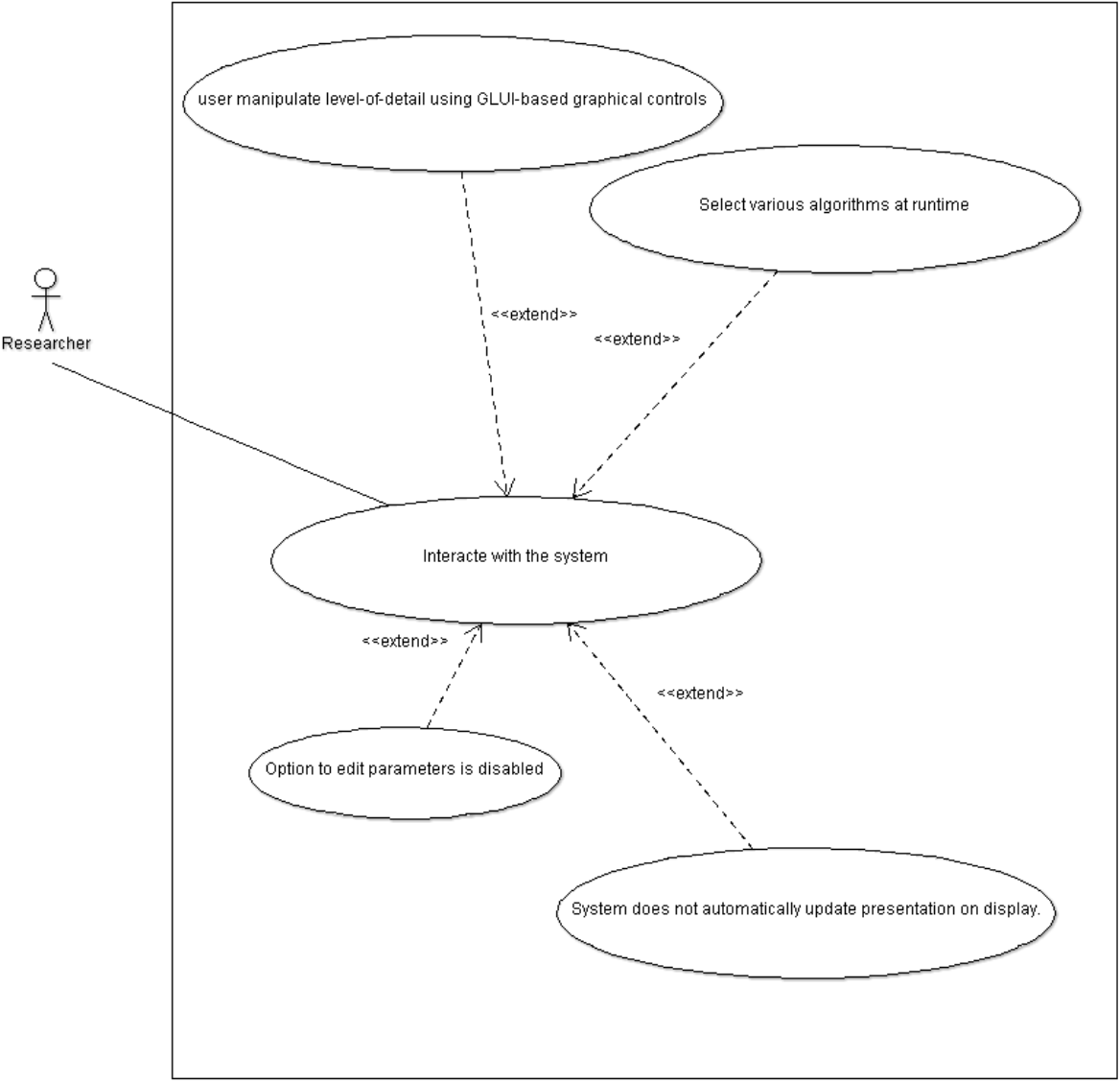

**Requirement 3:**

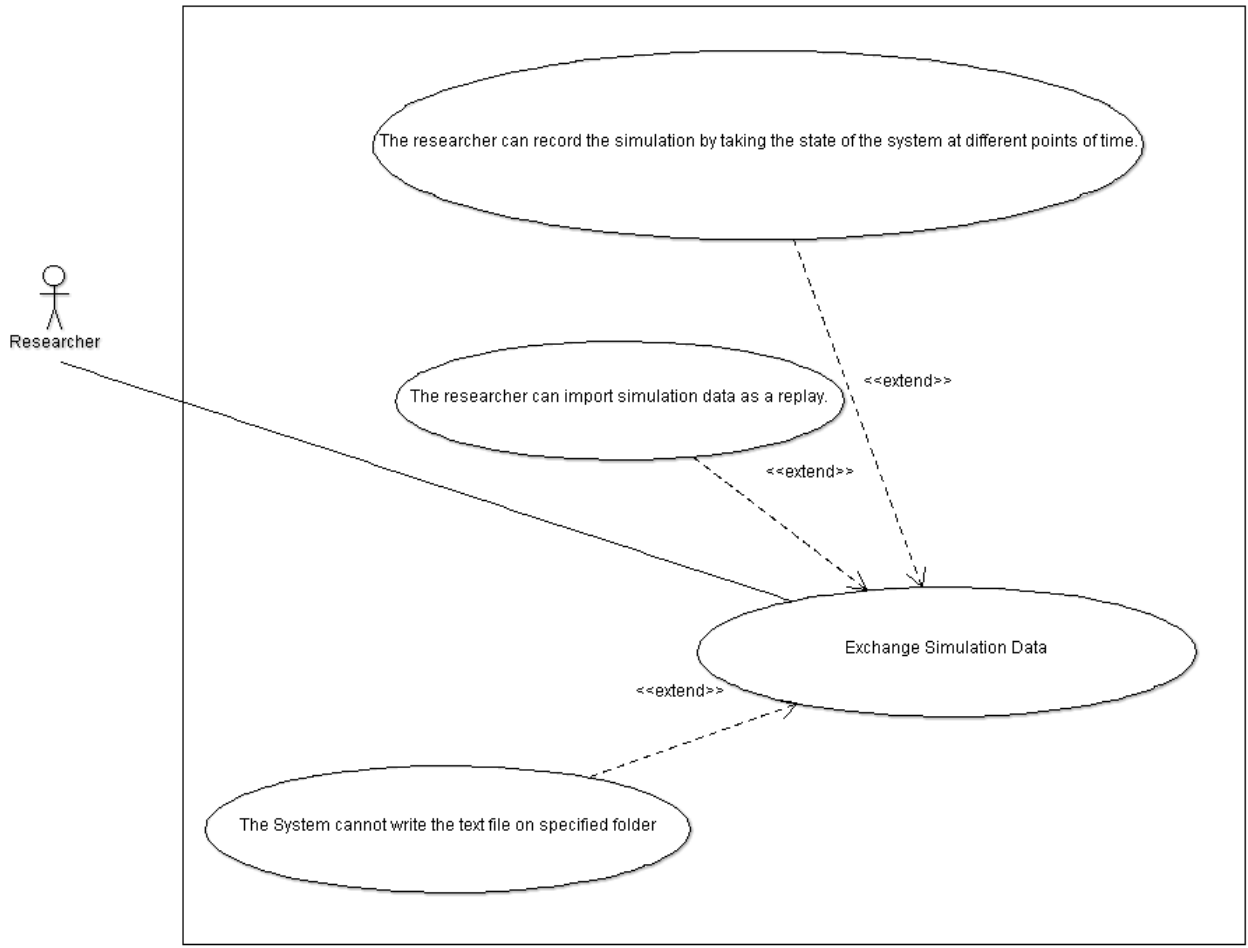

Concordia University

# Supplementary Specification and Glossary

SOEN 6481

**Prepared by:**

Klajdi Karanxha **|** 6173780
Herman Sonfack **|** 5986052
Gustavo Pereira **|** 6273033
Kapies Vallipuram **|** 9346287

## Revision History

| Date | Revision # | Description | Author(s) |
|---|---|---|---|
| 01-03-2013 | 1.0 | Supplementary Specification and Glossary | Klajdi Karanxha, Herman Sonfack, Kapies Vallipuram, Gustavo Pereira |
| 02-04-2013 | 2.0 | Passive voice, weak phrases optimization | Klajdi Karanxha, Herman Sonfack, Kapies Vallipuram, Gustavo Pereira |
| 17-04-2013 | 3.0 | Passive voice, weak phrases optimization | Klajdi Karanxha, Herman Sonfack, Kapies Vallipuram, Gustavo Pereira |

# Table of Contents

# 1. Introduction

Supplementary systems specifications for the requirements are described along the following document. In essence this document helps better identify and elaborate supplementary requirements that are not obvious or properly described in the use cases. It also elaborates on system requirements previously described in the use cases. The supplementary specifications focuses on standards and quality attributes of non-functional requirements and also on context and environment requirements. It will also felicitate legal and regulatory requirements and discuss other system environment related issues.

## 1.1. Purpose

The purpose of this document is to identify the non-functional requirements of the Softbody Simulation System and focus on its specifics and details.

## 1.2. Scope

The scope of this supplementary specification is to capture, analyze and elaborate non-functional requirements. It is mainly focused to the system's requirements for the Softbody Simulation System as our main project.

## 1.3. Definitions, Acronyms and Abbreviations

*Please consult* Chapter 14 - Glossary*.*

## 1.4. References

Please consult References Section at the end of the document.

## 1.5. Overview

The following sections contain information regarding the corresponding non-functional and functional requirements of the system. They describe how the requirements have been mapped to system functionality, what are the goals that the system should achieve and how to develop conform to non-functional requirements established.

## 2. Functionality

The section will provide details about system features (organized by feature) as per use case specifications. In addition, several functionalities may need a more detailed analysis regarding their usage and may be not covered entirely in use cases.

### 2.1. Choosing between multiple algorithms at run-time

The system user is able to choose between multiple algorithms on run time.

Related Feature (s): Visual Performance Analysis

### 2.2. Simulation data exporting and serialization

The system user is able to serialize exported data on external device.

Related Feature(s):     Simulation Performance Report; Rendering Performance Report.

### 2.3. Multiple Input devices

The system user is able to attach a hard-body device to act as sensor and interchange signals with software.

Related Feature(s): Interactivity through haptic device

### 2.4. Simulation alteration at runtime

The system user is allowed to use Graphic tools to alter the simulation at runtime and manipulate detail level.

Related Feature(s): User I/O LOD GUI

# 3. Usability

In this section we will discuss system usability. The system should be easy to use and understandable by the stakeholders. It also should be easy to use by system users that are already familiar with Softbody Simulation System's framework.

## 3.1. Training

System users should be able to use the system as fast as they can. The training process shall be comprehensive and users shall be provided with training prior of using the system and as they go. Initial training will increase usability standards of the system.

## 3.2. User Interface

System interface shall be compliant to standard UI accessibility features and follow a cognitive design which allows users for better interaction with the system. It also shall follow library inherited UI standards if any.

## 3.3. Daily Usage

System shall be available and reliable once running. Its daily performance shall remain steady and never degrade over time.

## 3.4. Mean Time between Failures

System shall be reliable. In event of a fault, error or misconfiguration, fixing the issue in addition to a simple system restart shall resolve the problem. Time between failures shall be no longer than system restart time (application re-launches).

## 4. Reliability

The system has to be very reliable as it used in research for computer graphic simulation.

### 4.1. Availability

The system is available 100% of the time for the users. The system shall be operational 24 hours a day for 365 days a year less the downtime for maintenance preferably on weekends.

### 4.2. Mean Time between Failures (MTBF)

One requirement is to implement statistics gathering for various real-time performance metrics for Simulation and Rendering. So the system used to perform real-time simulation hence it must not fail more than once a year.

### 4.3. Mean Time to Repair (MTTR)

Not specified but we can presume that the system cannot stay offline for more than 8 hours and any system error shall be addressed immediately by the development team.

### 4.4. Accuracy

Not specified

### 4.5. Maximum bug or defect rate

Not specified.

### 4.6. Bugs or defect rate

Not specified

# 5. Performance

## 5.1. Response Time

As per UC1 the user may alter simulation at run-time via GLUI-based graphical controls so the system answer shall be in a short time to the user.

## 5.2. Throughput

Researchers use the system for simulation and they can alter it at run-time. That implies a possible high number of transactions per second.

## 5.3. Capacity

Not specify but we can presume that the system is compiled and run locally on the user machine.

## 5.4. Resource

As per UC2, user shall select various algorithms at runtime for comparative studies on various issues. Enough memory must be in place to accommodate that.

# 6. Supportability

## 6.1. Coding Standards

The system compiles and runs on the Microsoft Windows platform and accommodates other platform, such as Linux and Mac OS X. The system is written in C++.  A Plan shall be there for deployment and building under different platforms and build system. E.g. Linux with Makefiles and autoconf, or Mac OS X with Xcode or also Makefiles. The standard for each language is not specified

## 6.2. Class Libraries

 The Libraries include the OpenGL along with the drivers for OLSL support, GLUI or even less obvious CUGL. Other plans include DirectX and HD support. The system exports and generated as a library or a collection of library and API's.

## 6.3. Naming convention

The exported API and global shall be constrained in the system's own namespace to avoid clashes with external applications during linking.

# 7. Design Constraints

## 7.1. Software Language Used

Portability of the source code (at minimum) and plan for deployment and building under different platforms and build systems, e.g. Linux with Makefiles and autoconf, or Mac.

## 7.2. Development tools

API hooks should be always provided for plug-in architectures for all algorithm where there could be more than one instance, such as integration, subdivision, collision detection in order to simply the integration effort with other projects and other programming languages.

Also the exported API and global should be constrained in the system's own namespace to avoid clashes with external applications during linking.

The main algorithm adjusts with the extension of subclasses applications so it is less rigid.

## 7.3. Class Libraries

The Libraries include the OpenGL along with the drivers for OLSL support, GLUI or even less obvious CUGL. Other plans include DirectX and HD support. The system can export and generated as a library or a collection of library and API's.

# 8. Online User Documentation and Help System Requirements

A plan for academic value of teaching and learning computer graphics and physical based simulations, by structuring the code, comments, and documentation per consistent naming and coding convention and the APU

# 9. Purchased Components

No licensing has been specified. No required purchased components to be used either but the system shall cover interoperability with different games and rendering engines, potentially distributed, extended simulation systems and the use of the system as a library

# 10. Interfaces

## 10.1. User Interfaces

The user interfaces of the system should be designed using GLUI, where researchers can test different physical based phenomena at real-time, manipulating parameters and the level-of-detail (LOD) to reach the usability level desired.

## 10.2. Hardware Interfaces

The user usually interacts with the system via mouse and keyboard. This interaction can be enhanced with re-active controls (i.e. haptic devices) to provide force feedback. Those devices operated with support of device drivers (provided by manufacturer to corresponding OS) and the proper API hook for the desire functionality.

## 10.3. Software Interfaces

External software interfaces may operate the system using same API hooks provided for external hardware interfaces. These interactions are restricted to same functionalities of user I/O GUI for manipulating parameters.

## 10.4. Communications Interfaces

The system is designed as a standalone application, without expectation to handle or manage any external communication. Eventually, it can be done by external hardware/software described above.

# 11. Licensing Requirements

This system is a collaborative open-source project for academic study purpose. It complies with Concordia University's Code of Conduct and it's under the provisions of a Software Assignment submission. It may be license with a Creative Common 3.0 license, attributes BY (by attribution) and SA (share alike) [1].

## 12. Legal, Copyright and Other Notices

The Softbody Simulation System are copyright their respective owners. All rights reserved. Permission to use all or part of this work for personal or classroom use is granted without fee provided that copies are not made or distributed for profit or commercial advantage and that copies bear this notice and the full citation. Copyright content must be submitted to respective owners prior to commercial use. Organizations involved in this project may be consulted [2][3][4].

The Softbody Simulation System is intended to academic study only. Researchers involved on this project are not liable for any losses or injuries caused by misuse of this system or the results of this system. All users of this system hereby indemnify any and all responsibilities from the researchers as a result of using the system.

# 13. Applicable Standards

The Softbody Simulation System must follow, meet, and compliant with all ISO/IEC/IEEE standards for Information Technology and Software Engineering, more specifically the following standards:

- ISO/IEC 25010:2011 Systems and software Quality Requirements and Evaluation (SQuaRE) - ⁭System and software quality models [5]
- ISO/IEC TR 25060:2010 Systems and software Quality Requirements and Evaluation (SQuaRE) - ⁭Common Industry Format (CIF) for usability: General framework for usability-related information [6]
- ISO/IEC 26514:2008 Requirements for designers and developers of user documentation [7]
- ISO/IEC 90003:2004 Guidelines for the application of ISO 9001:2000 to computer software [8]

# 14. Glossary

**API** – Application Program Interface is a protocol intended to be used as an interface by software components to communicate with each other. [9]

**APU** - Accelerated processing unit

**CUGL -** Concordia University Graphics Library

**GLUI** - OpenGL User Interface Library [14]

**Mean time between failures (MTBF) -** Is the predicted elapsed time between inherent failures of a system during operation [12]

**Mean time to repair** (**MTTR**) is a basic measure of the maintainability of repairable items. It represents the average time required to repair a failed component or device. [13]

**Open Source** – Is a software development or broader philosophy which promotes free distribution of software design and implementation.

**OpenGL** - Open Graphics Library

**Performance** - the execution of an action [11]

**Source Code** – In computer science, source code is any collection of computer instructions (possibly with comments) written using some human-readable computer language, usually as text. The source code of a program is specially designed to facilitate the work of computer programmers, [10]

**UC** – Use Case

**UI** – User Interface

Concordia University

# Test Cases

SOEN 6481

**Prepared by:**

Klajdi Karanxha **|** 6173780
Herman Sonfack **|** 5986052
Gustavo Pereira **|** 6273033
Kapies Vallipuram **|** 9346287

# Revision History

| Date | Revision # | Description | Author(s) |
|---|---|---|---|
| 17-04-2013 | 1.0 | Initial Document of a fully described Test Case, using Pre and Post conditions and settings. | Klajdi Karanxha, Herman Sonfack, Kapies Vallipuram, Gustavo Pereira |

# 1. INTRODUCTION

The following test cases target the two fully-dressed use cases of the system, their main and alternative scenarios. Each of these Use case scenarios leads to several test scenarios each of those scenarios composed of several test cases with their own plan and execution.

## 2. TEST CASES

The test cases for each test scenario of the use case are described in the tables that follow.

## 2.1. Alter Simulation at Run-Time Via GLUI-Based Graphical Controls

This is one of the main use cases described for the system. This use case has a success scenario where the user manages to change simulation via the controls and the failure scenarios, where the user fails to alter the simulation at runtime.

### 2.1.1. Main Success Scenario

| Test Case ID | 2.1.1 |
|---|---|
| Title | Successful simulation alternation. |
| Requirement | Appropriate set of user permissions. Multiple monitors to display all representations possible. |
| Type | Regular. |
| Settings | Simulation software is running. |
| Preconditions | Simulation software is running and the view is set up appropriately. |
| Description | Researcher access graphical control panel. Edits simulation parameters. Confirms parameter changes. |
| Expected Results | Updates take effect while simulation continues running. |

## 2.1.2. Alternate Scenarios

| Test Case ID | 2.1.2 |
|---|---|
| Title | Wrong permissions. |
| Requirement | Researcher has permissions to use the tool.<br>Multiple monitors to display all representations possible. |
| Type | Regular. |
| Settings | Simulation software is running. |
| Preconditions | n/a |
| Description | Researcher access graphical control panel.<br>Attempts to edit parameters. |
| Expected Results | A warning message reminds the user about his/her permissions. |

| Test Case ID | 2.1.3 |
|---|---|
| Title | Failure to reflect parameter change |
| Requirement | Multiple monitors to display all representations possible. |

| Type | Regular. |
|---|---|
| Settings | Simulation software is running. |
| Preconditions | n/a |
| Description | Researcher edits parameters. Saves configuration. Closes the software. Re-loads with new configuration. |
| Expected Results | Simulation doesn't refresh even after manual change. |

## 2.2. Select Various Algorithms at Runtime for Comparative Studies on Various Issues

This use case is also part of the main use cases of the system. It allows the researcher to select and choose between various computations algorithms at runtime.

### 2.2.1. Main Success Scenario

| Test Case ID | 2.2.1 |
|---|---|
| Title | Change Algorithm at Runtime |
| Requirement | Multiple monitors to display all representations. |
| Type | Regular. |
| Settings | Researcher has access to algorithm selection. |
| Preconditions | Collection of algorithms has been |

|  | pre-loaded in the system. |
|---|---|
| Description | Researcher loads menu.<br>Selects specific algorithm from the list.<br>Saves changes. |
| Expected Results | Simulation runs under the new algorithm. |

### 2.2.2. Alternate Scenarios

| Test Case ID | 2.2.2 |
|---|---|
| Title | Empty algorithm selection |
| Requirement | Multiple monitors to display all representations. |
| Type | Regular. |
| Settings | Researcher has access to algorithm selection. |
| Preconditions | n/a |
| Description | Researcher attempts to load a new algorithm. |
| Expected Results | List of algorithms is empty. |

| Test Case ID | 2.2.3 |
|---|---|
| Title | Invalid parameter load |

| Requirement | Multiple monitors to display all representations. |
|---|---|
| Type | Regular. |
| Settings | Researcher has access to algorithm selection. |
| Preconditions | n/a |
| Description | Researcher loads a new algorithm from the menu. Manually refresh display. |
| Expected Results | Simulation doesn't refresh, the new algorithm is not loaded. An error message warns the user. |

| Test Case ID | 2.2.4 |
|---|---|
| Title | Discouraged selection of algorithm |
| Requirement | Multiple monitors to display all representations. |
| Type | Sanity plus. |
| Settings | Researcher has access to algorithm selection |
| Preconditions | The system is running under optimal algorithm |

|  | performance. |
|---|---|
| Description | Researcher selects a "less-optimal" or inappropriate algorithm from the menu. Saves the selection |
| Expected Results | System performance degradation. A message shall indicate recommended performance. |

Concordia University

# Traceability Document

SOEN 6481

**Prepared by:**

Klajdi Karanxha **|** 6173780
Herman Sonfack **|** 5986052
Gustavo Pereira **|** 6273033
Kapies Vallipuram **|** 9346287

# Revision History

| Date | Revision # | Description | Author(s) |
|---|---|---|---|
| 17-04-2013 | 1.0 | Final Document of traceability | Klajdi Karanxha, Herman Sonfack, Kapies Vallipuram, Gustavo Pereira |

# 1. INTRODUCTION

Traceability in our project will help maintain a high level of software quality and reliability. As well, it may add value to the development process by increasing the degree of the relationship between different project artefacts.

# 2. TRACEABILITY

Traceability links are analyzed in implementation and testing areas and are based on the requirements specifications artefact.

## 2.1. Needs to Features

As stated, user needs map to system features which are prioritized. Features based on their priority are decided in which iteration to be implemented. In order to preserve consistency and prepare for later change, traceability comes handy.

In order to implement the traceability matrix and taking the consideration the dimensions of the matrix, we will assign <u>numbers to the features</u> and <u>letters to the user needs</u> and map them in the matrix.

### 2.1.1. User Needs

List of User Needs (from vision document and project description):

A. Implement statistics gathering for various real-time performance metrics for Simulation and Rendering.

B. Allow the user to alter the simulation at run-time via GLUI-based graphical controls to manipulate the level-of-detail (LOD) and simulation parameter.

C. Run-time selection for comparative studies on any aspect of visual realism to run-time performance and memory usage.

D. Lighting and texture mapping techniques.

E. Allow objects be "attached" to "hardbody" objects providing points of attachment.

F. Allow alteration of the Archimedean-based graphs and different types of them than a octahedron as a run-time LOD parameter.

G. Interactivity through haptic devices.

H. Allow the state dump and reload functionality at any given point in time enabling to reproduce a simulation from some point in time.

    I.   Allow for stereoscopic effects.

### 2.1.2. List of Features

The list of the features that are planned for development:

1. Simulation Performance Report.
2. Rendering Performance Report.
3. User I/O LOD GUI.
4. Visual Performance Analysis.
5. Lighting and texture.
6. Attachment.
7. Archimedean-based graphs Alteration.
8. Device Control.
9. Run Time Watch Point.
10. Stereoscopic effects.

## 2.1.3. Traceability Matrix

| Need/Feature | 1 | 2 | 3 | 4 | 5 | 6 | 7 | 8 | 9 | 10 |
|---|---|---|---|---|---|---|---|---|---|---|
| A | x | x | | | | | | | | |
| B | | | x | | | | | | | |
| C | | | | x | | | | | | |
| D | | | | | x | | | | | |
| E | | | | | | x | | | | |
| F | | | | | | | x | | | |
| G | | | | | | | | x | | |
| H | | | | | | | | | x | |
| I | | | | | | | | | | x |

## 2.2. Tracing Features to Use Cases

Every use case discussed involves the main scenario of how a feature would be used by system actors. Below there is a mapping between features and use cases.

### 2.2.1. List of Use Cases

The list of the use cases is as follows:

- 1.1 – Execute multiple algorithms / Select various algorithms.
- 1.2 – Export simulation data.
- 1.3 – Attach soft-body like objects to hard-body objects.
- 1.4 – Alter simulation at run-time using graphic controls.

The features list is numbered previously.

### 2.2.2. Traceability Matrix

| Feature/Use Case | 1.1 | 1.2 | 1.3 | 1.4 |
|---|---|---|---|---|
| 1 | | x | | |
| 2 | | x | | |
| 3 | | | | x |
| 4 | | x | | x |
| 5 | | | | |
| 6 | | | x | |
| 7 | x | | | |
| 8 | | | x | |
| 9 | | | | x |
| 10 | | | | |

## 2.3. Features and Supplementary Requirements

All the supplementary requirements apply to some or many features; in this section we provide a mapping for traceability purposes.

### 2.3.1. List of Supplementary Requirements

The most important non-functional requirements to be met by the system are listed below:

**A.** Usability

**B.** Reliability

**C.** Performance

**D.** Supportability

### 2.3.2. Traceability Matrix

| Feature/Supplementary Reqs. | A | B | C | D |
|---|---|---|---|---|
| 1 | x |   | x |   |
| 2 | x |   | x |   |
| 3 | x |   |   |   |
| 4 | x |   | x |   |
| 5 | x |   |   |   |
| 6 | x |   |   |   |
| 7 | x |   | x | x |
| 8 |   | x |   | x |
| 9 |   | x |   |   |
| 10 |   |   | x |   |

## 2.4. Use Cases to Test Cases

Use cases of the system comprise a main scenario and in some cases alternative or failure scenarios. Based on use cases we can have different test case scenarios leading to different set of results for each use case.

## 2.4.1. Use Case and Scenarios

The following matrix maps the use cases to different testing scenarios that can be derived from their flow. For the purpose of this project we only discuss two fully dressed use case scenarios. Then every scenario is mapped to a specific test case that shows the settings, pre-condition, post-condition and description of every possible test scenario that can be performed or that specific use case flow. The test cases are listed as below:

| Use Case | Scenario Number | Originating Flow | Alternate Flow | Next Alternate | Test Case |
|---|---|---|---|---|---|
| **1.1** | 1 | Change to new algorithm. | | | 2.2.1 |
| | 2 | Change to new algorithm. | Empty List. | | 2.2.2 |
| | 3 | Change to new algorithm. | Failed Update. | | 2.2.3 |
| **1.4** | 1 | Simulation runs under new parameters. | | | 2.1.1 |
| | 2 | Simulation runs under new parameters. | GLUI-based flow. Simulation runs under new parameters. | | 2.1.2 (note: different test case settings for GLUI) |
| | 3 | Simulation runs under new parameters. | Disabled option to edit parameters. | | 2.1.2 |
| | 4 | Simulation runs under new parameters. | Failed update on display. | | 2.1.3 |

Concordia University

# Activity Diagram

SOEN 6481

**Prepared by:**

Klajdi Karanxha **|** 6173780
Herman Sonfack **|** 5986052
Gustavo Pereira **|** 6273033
Kapies Vallipuram **|** 9346287

# Revision History

| Date | Revision # | Description | Author(s) |
|---|---|---|---|
| 17-04-2013 | 1.0 | Final Revision of Activity Diagram | Klajdi Karanxha, Herman Sonfack, Kapies Vallipuram, Gustavo Pereira |

# 1. Activity Diagram UC 2.0

Activity Diagram UC 2.0

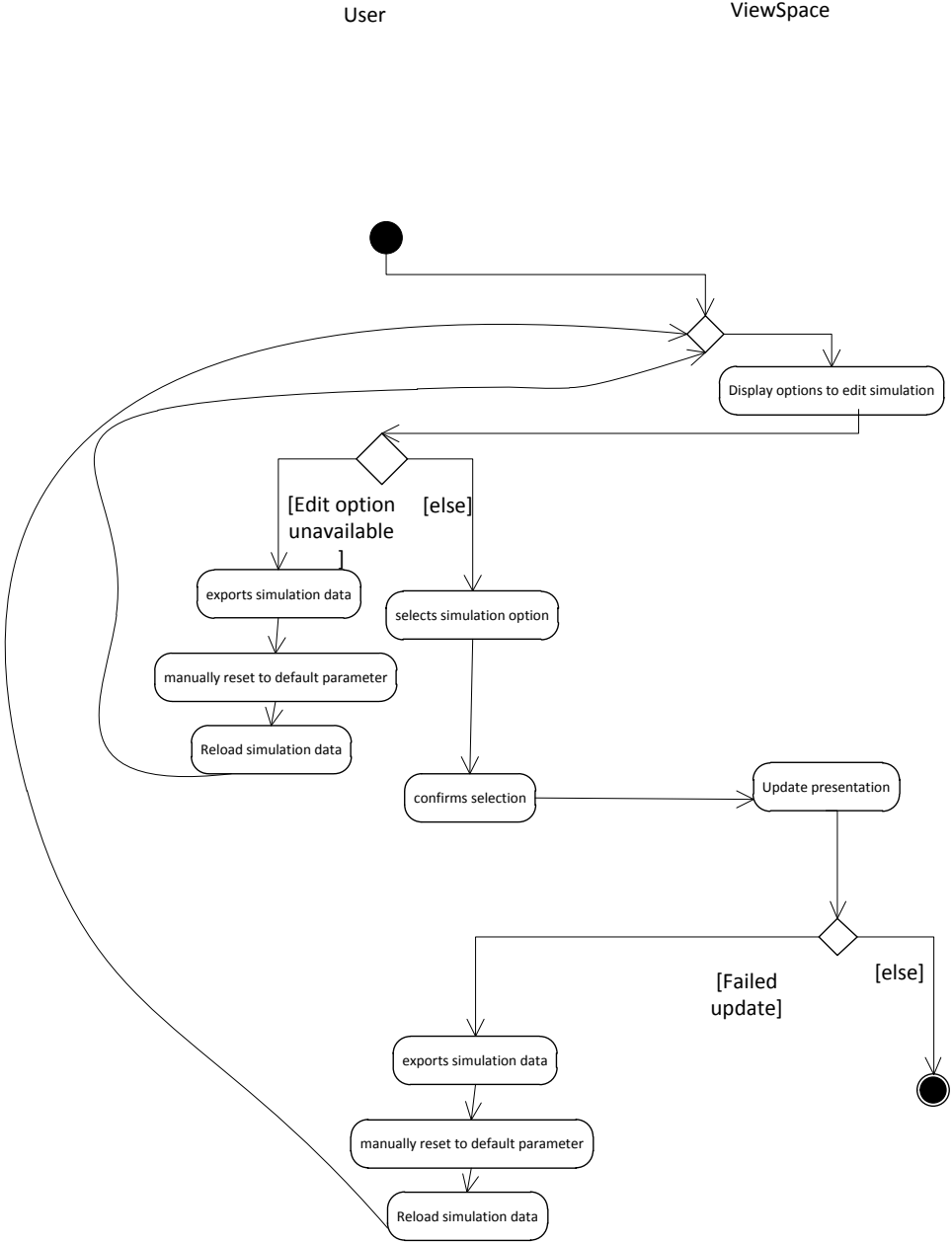

## 2. Activity Diagram UC 3.0

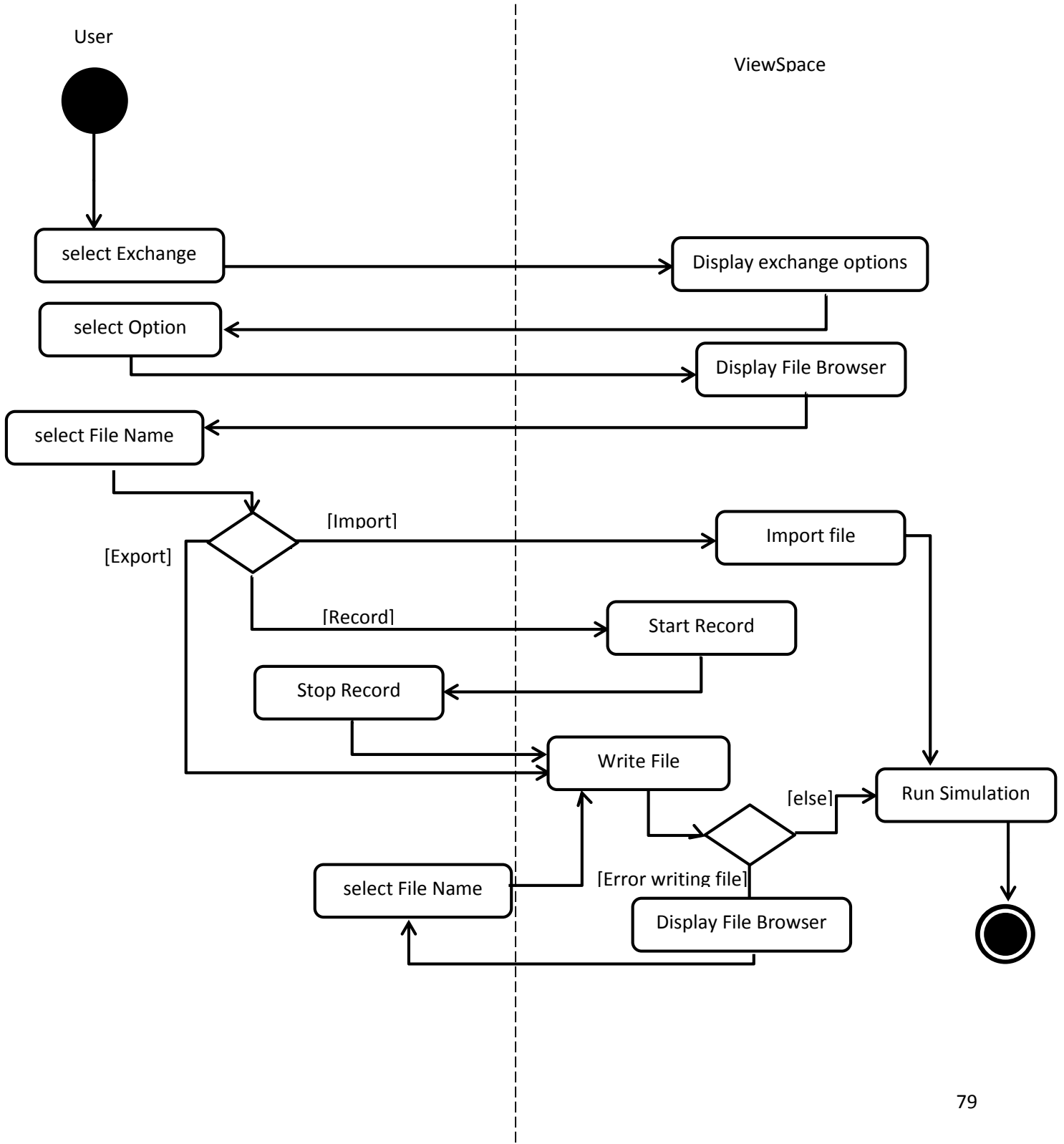

Concordia University

# Domain Model

SOEN 6481

**Prepared by:**

Klajdi Karanxha **|** 6173780
Herman Sonfack **|** 5986052
Gustavo Pereira **|** 6273033
Kapies Vallipuram **|** 9346287

## Revision History

| Date | Revision # | Description | Author(s) |
|---|---|---|---|
| 17-04-2013 | 1.0 | Final revision of Domain Model | Klajdi Karanxha, Herman Sonfack, Kapies Vallipuram, Gustavo Pereira |

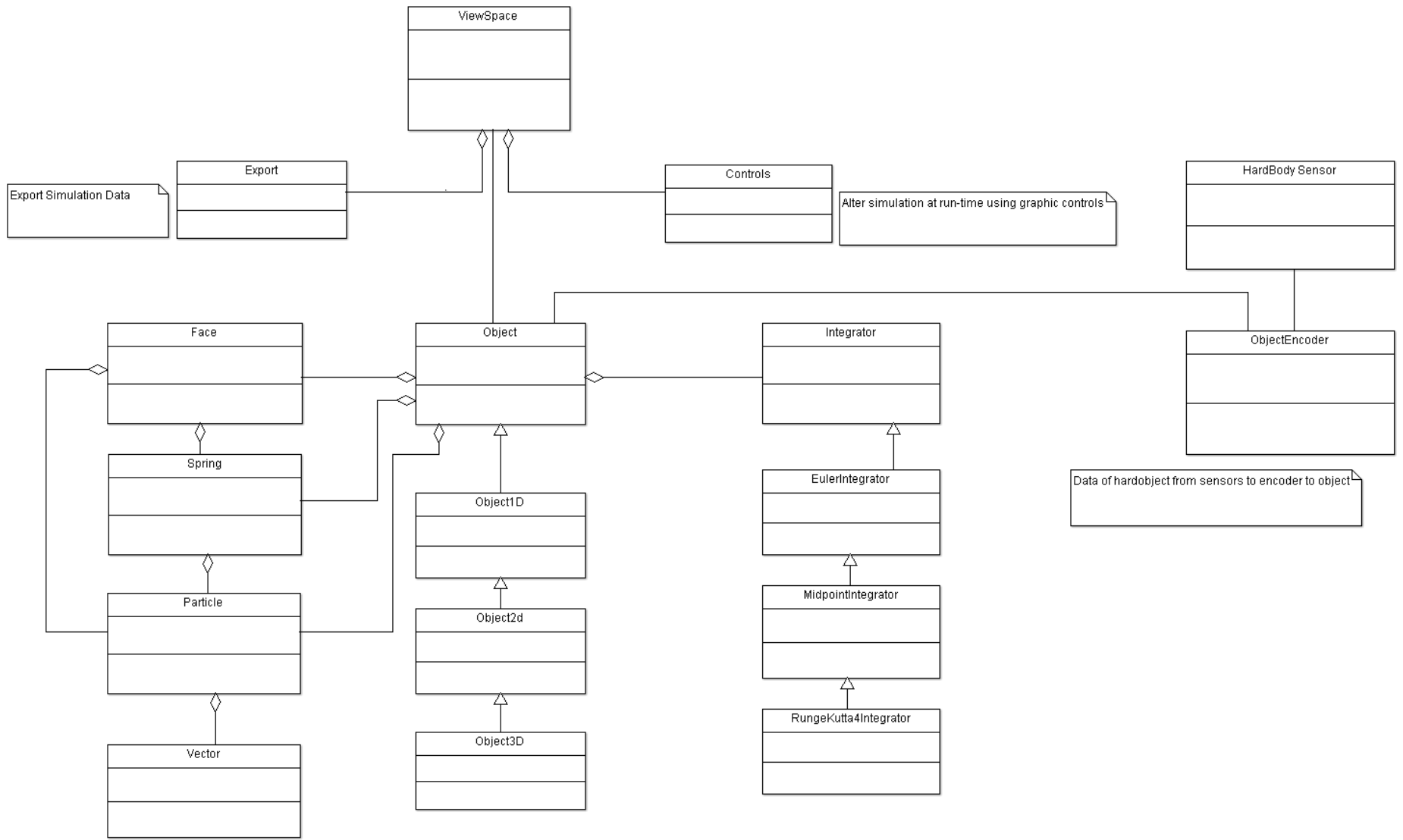

Concordia University

# SSD – System Sequence Diagram
SOEN 6481

# Revision History

| Date | Revision # | Description | Author(s) |
|---|---|---|---|
| 17-04-2013 | 1.0 | Final Document of SSD – System Sequence Diagram | Klajdi Karanxha, Herman Sonfack, Kapies Vallipuram, Gustavo Pereira |

## UC 2.0: Interact with the system (Main Success Scenario)

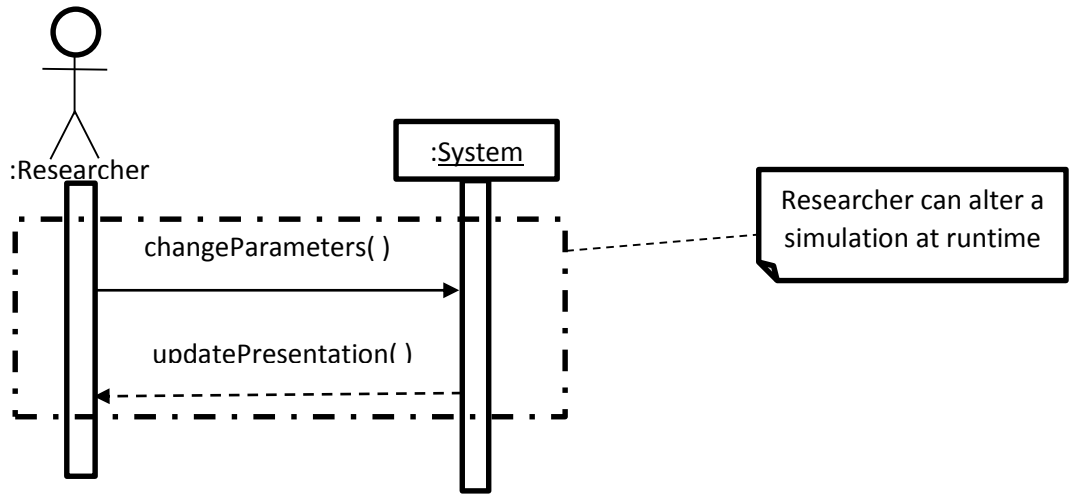

## UC 3.0: Exchange Simulation Data (Main Success Scenario)

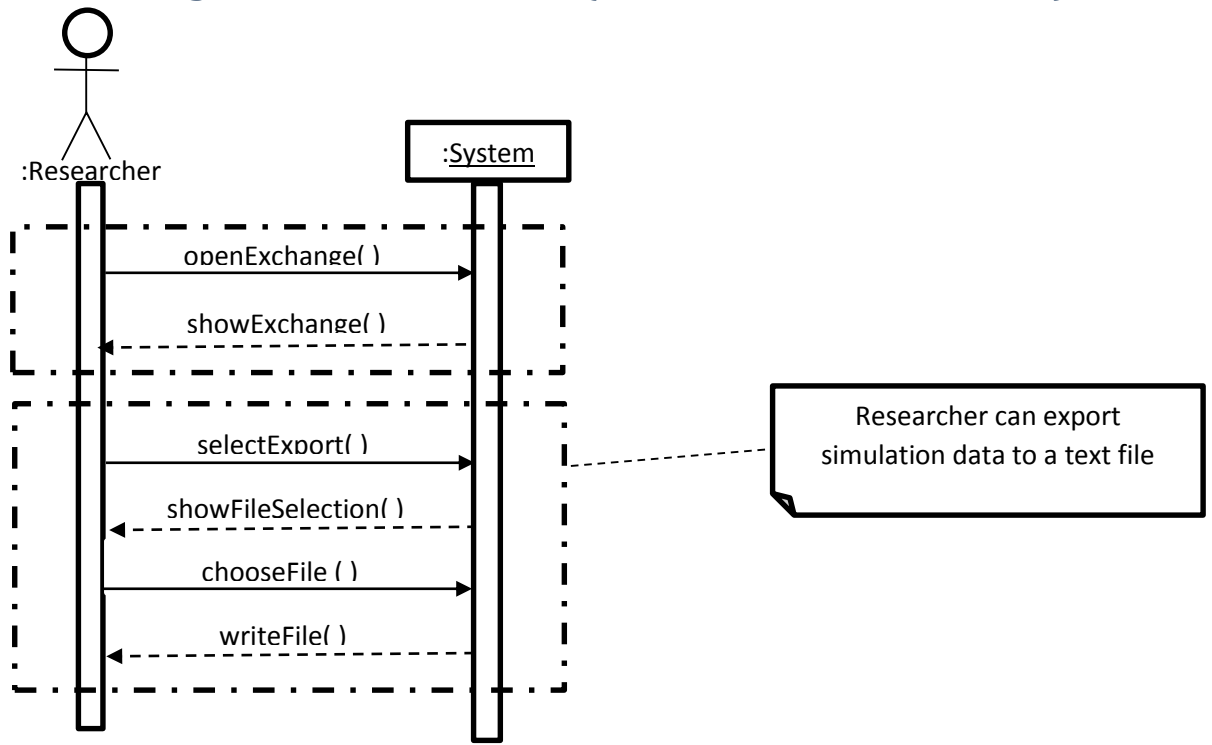